\documentclass{article}
\usepackage{ijcai24}

\usepackage{times}
\usepackage{soul}
\usepackage{url}
\usepackage[hidelinks]{hyperref}
\usepackage[utf8]{inputenc}
\usepackage[small]{caption}
\usepackage{graphicx}
\usepackage{amsmath}
\usepackage{amsthm}
\usepackage{booktabs}
\usepackage{algorithm}
\usepackage{algorithmic}
\usepackage[switch]{lineno}

\usepackage{etoolbox}
\robustify\bfseries
\usepackage{graphicx}
\usepackage{graphics}
\usepackage{amsmath,amssymb,graphicx}
\usepackage{tabularx}
\usepackage{wasysym}
\usepackage{color,soul}
\usepackage{balance}
\usepackage{breqn}
\usepackage{makecell}
\usepackage{booktabs}
\usepackage{multirow}
\usepackage{siunitx}
\usepackage{xcolor}

\usepackage{hyperref}
\def\H{{\mathsf H}}
\def\T{{\mathsf T}}
\def\CC{{\mathbb C}}
\def\RR{{\mathbb R}}
\usepackage{arydshln}
\usepackage{enumitem}
\usepackage{float}

\usepackage{amssymb}%
\usepackage{pifont}%
\newcommand{\cmark}{\ding{51}}%
\newcommand{\xmark}{\ding{55}}%

\newcommand\mydots{\hbox to 1em{.\hss.\hss.}}

\title{Cross-Talk Reduction}

\author{
Zhong-Qiu Wang$^1$
\and
Anurag Kumar$^2$
\And
Shinji Watanabe$^3$\\
\affiliations
$^1$Southern University of Science and Technology, China\\
$^2$Meta Reality Labs Research, USA\\
$^3$Carnegie Mellon University, USA\\
\emails
wang.zhongqiu41@gmail.com
}

\begin{document}

\maketitle

\begin{abstract}
While far-field multi-talker mixtures are recorded, each speaker can wear a close-talk microphone so that close-talk mixtures can be recorded at the same time.
Although each close-talk mixture has a high signal-to-noise ratio (SNR) of the wearer, it has a very limited range of applications, as it also contains significant cross-talk speech by other speakers and is not clean enough.
In this context, we propose a novel task named \textit{cross-talk reduction} (CTR) which aims at reducing cross-talk speech, and a novel solution named CTRnet which is based on unsupervised or weakly-supervised neural speech separation.
In unsupervised CTRnet, close-talk and far-field mixtures are stacked as input for a DNN to estimate the close-talk speech of each speaker.
It is trained in an unsupervised, discriminative way such that the DNN estimate for each speaker can be linearly filtered to cancel out the speaker's cross-talk speech captured at other microphones. 
In weakly-supervised CTRnet, we assume the availability of each speaker's activity timestamps during training, and leverage them to improve the training of unsupervised CTRnet.
Evaluation results on a simulated two-speaker CTR task and on a real-recorded conversational speech separation and recognition task show the effectiveness and potential of CTRnet.
\end{abstract}

\section{Introduction}

\begin{figure}
  \begin{center}
  \includegraphics[width=0.42\textwidth]{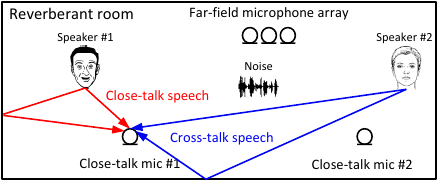}
  \end{center}
  \caption{
  Task illustration.
  Best viewed in color.
  }\label{physical_model_figure}
\end{figure}

While far-field mixtures of multiple speakers are recorded, the close-talk mixture of each speaker is often recorded at the same time, by placing a microphone close to each target
speaker (e.g., in the AMI \cite{McCowan2006}, CHiME \cite{Barker2018CHiME5}, AliMeeting \cite{Yu2022M2MeT}, and MISP \cite{Wang2023MISP} setup).\footnote{Close-talk mixtures are almost always recorded in conversational speech separation and recognition datasets, since it is much easier for humans to annotate transcriptions and speaker-activity timestamps based on close-talk mixtures than far-field mixtures.
}
See Fig. \ref{physical_model_figure} for an illustration.
Each close-talk mixture consists of 
close-talk speech, cross-talk speech and environmental noises, all of which are reverberant.
Inside close-talk mixtures, the close-talk speech usually has a strong energy, and inside close-talk speech, the direct-path signal of the target speaker is typically much stronger than reverberation.
However, besides close-talk speech, close-talk mixtures often contain significant cross-talk speech produced by other speakers, especially if the other speakers, while talking, are spatially close to the target speaker.
The contamination of cross-talk speech dramatically limits the application range of close-talk mixtures.
For example, they are seldomly exploited as supervisions for training far-field speech separation models or leveraged as reference signals for evaluating speech separation models.
It is a common perception that they are not suitable for these purposes as they are not clean enough \cite{Barker2018CHiME5,HaebUmbach2019SPM,Wisdom2020MixIT,Watanabe2020CHiME6,Sivaraman2022,Aralikatti2022RAS,cornell23_chime,Leglaive2023CHiME7UDASE}.

This paper aims at reducing cross-talk speech and separating close-talk speech in close-talk mixtures.
We name this task \textit{cross-talk reduction}.
Solving this task could enable many applications.
For example, we can (a) leverage separated close-talk speech as pseudo-labels for training supervised far-field separation models; (b) use it as reference signals to evaluate the separation results of far-field separation models; and (c) present it rather than close-talk mixtures to annotators to reduce their annotation efforts.

One possible approach for cross-talk reduction is supervised speech separation \cite{WDLreview,Yin2020,Jenrungrot2020,Tang2020,Nachmani2020,Rixen2022,Wang2022GridNetjournal}, where synthetic pairs of clean close-talk speech and close-talk (and far-field) mixtures are first synthesized via room simulation and then used to train supervised learning based models to predict the clean close-talk speech from its paired mixtures.
The trained models, however, are known to suffer from severe generalization issues, as the simulated data used for training is often and inevitably mismatched with real-recorded test data \cite{WDLreview,Pandey2020CrossCorpus,Tzinis2020AudioScope,Zhang2021ClosingGap,Tzinis2022REMIXT,Tzinis2022AudioScopeV2,Leglaive2023CHiME7UDASE}.

Given only paired real-recorded far-field and close-talk mixtures, we propose to tackle cross-talk reduction by training unsupervised speech separation models directly on the real-recorded mixtures.
This way, we can avoid the effort of data simulation, and the issues incurred when using simulated data.
We point out that many existing unsupervised separation algorithms could be employed for cross-talk reduction.
In this paper, we design a potentially better unsupervised algorithm by leveraging the fact that, for each speaker, the close-talk speech in close-talk mixtures has a strong energy, which can provide an informative cue about what the target speech is.
Our idea is that we can (a) roughly separate the close-talk speech from each close-talk mixture (this could be achieved with a reasonable performance since the SNR of close-talk speech is usually high);
and (b) reverberate the separated close-talk speech of each speaker via linear filtering to identify, and then cancel out, the speaker's cross-talk speech captured by the close-talk microphones of the other speakers.
Based on this idea, we propose CTRnet, where, during training, close-talk and far-field mixtures are stacked as input for a deep neural network (DNN) to first produce an estimate for each close-talk speech, and then the cross-talk speech in each close-talk mixture is identified (and cancelled out) by linearly filtering the DNN estimates via a linear prediction algorithm such that the filtering results of all the speakers can add up to the mixture.
This paper makes three major contributions:
\begin{itemize}[leftmargin=*,noitemsep,topsep=0pt]
\item We propose a novel task, cross-talk reduction.
\item We propose a novel solution, unsupervised CTRnet.
\item We further propose weakly-supervised CTRnet, where, during training, speaker-activity timestamps are leveraged as a weak supervision to improve unsupervised CTRnet.
\end{itemize}
Evaluation results on a simulated CTR task and on a real-recorded conversational speech separation and recognition task show the effectiveness and potential of CTRnet.
A sound demo is provided in the link below.\footnote{See {\url{https://zqwang7.github.io/demos/CTRnet_demo/index.html}}}

\section{Related Work}

To the best of our knowledge, we are the first studying cross-talk reduction.
There are tasks with similar names such as \textit{cross-talk cancellation} \cite{Bleil2023} which deals with sound field manipulation in spatial audio, and \textit{cross-talk suppression} \cite{Tripathi2022} in circuits design.
They are not related to our task and algorithms.

CTRnet builds upon the UNSSOR algorithm \cite{Wang2023UNSSOR}, leveraging UNSSOR's capability at unsupervised speech separation.
UNSSOR is designed to perform unsupervised separation based on far-field mixtures, assuming compact microphone arrays, while unsupervised CTRnet performs separation based on not only far-field but also close-talk mixtures, dealing with distributed-array cases for a different task: cross-talk reduction.
We further propose weakly-supervised CTRnet, which leverages speaker-activity timestamps as weak supervision and shows strong performance on real-recorded data.
In contrast, UNSSOR deals with unsupervised separation and is only validated on simulated data.

\section{Problem Formulation}

In a reverberant enclosure with $C$ speakers (each wearing a close-talk microphone) and a $P$-microphone far-field array (see Fig. \ref{physical_model_figure} for an illustration), each of the recorded closed-talk and far-field mixtures can be respectively formulated in the short-time Fourier transform (STFT) domain as follows:
\begin{align}
Y_{c}(t,f) &= \sum\nolimits_{c'=1}^C X_{c}(c',t,f) + \varepsilon_{c}(t,f), \label{physical_model_ct} \\
Y_p(t,f) &= \sum\nolimits_{c=1}^C X_p(c,t,f) + \varepsilon_p(t,f), \label{physical_model_ff}
\end{align}
where $t$ indexes $T$ frames, $f$ indexes $F$ frequencies, $c$ indexes $C$ speakers (and close-talk microphones), and $p$ indexes $P$ far-field microphones. 
$Y_{c}(t,f)$, $X_{c}(c',t,f)$ and $\varepsilon_{c}(t,f)$ in (\ref{physical_model_ct}) respectively denote the STFT coefficients of the close-talk mixture, reverberant image of speaker $c'$, and non-speech signals captured by the close-talk microphone of speaker $c$ at time $t$ and frequency $f$.
Notice that we use subscript $c$ to index the $C$ close-talk microphones.
Similarly, $Y_{p}(t,f)$, $X_p(c,t,f)$ and $\varepsilon_p(t,f)$ in (\ref{physical_model_ff}) respectively denote the STFT coefficients of the far-field mixture, reverberant image of speaker $c$, and non-speech signals captured at far-field microphone $p$.
In the rest of this paper, we refer to the corresponding spectrograms when dropping $p$, $c$, $t$ or $f$ in notations.
$\varepsilon$ is assumed as a weak and stationary noise term.

Based on the close-talk and far-field mixtures, we aim at estimating $X_{c}(c)$, the close-talk speech of each speaker $c$.\footnote{In $X_{c}(c)$, symbol $c$ in the subscript denotes the close-talk microphone of speaker $c$, and $c$ in the parenthesis indicates that the notation represents the reverberant image of speaker $c$.}
$X_{c}(c)$ is very clean and can be roughly viewed as the dry source signal.\footnote{Some reverberation of the source signal still exists in $X_c(c)$, but it is much weaker than the direct-path signal.
}
While speaker $c$ is speaking, $X_{c}(c)$ inside $Y_c$ is usually stronger than the cross-talk speech $X_{c}(c')$ by any other speaker $c'$ ($\neq c$).
With these understandings, we formulate the above physical models as (\ref{physical_model_ct_conv}) and (\ref{physical_model_ff_conv}), and formulate cross-talk reduction as a blind deconvolution problem in (\ref{ideal_loss}).

\begin{figure*}
  \centering  
  \includegraphics[width=17cm]{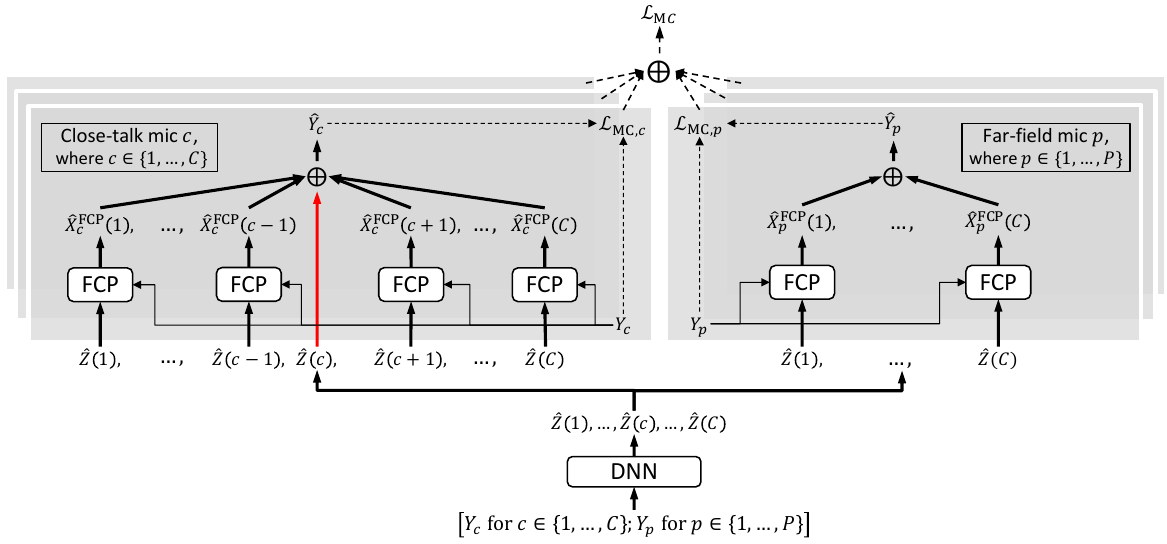}
  \caption{
   Illustration of unsupervised CTRnet (see first paragraph of Section \ref{proposed_algrithms_unsupervised} for detailed description).
  }
  \label{system_figure}
\end{figure*}

In detail, let $Z(c)=X_c(c)$ denote the close-talk speech of speaker $c$, we re-formulate (\ref{physical_model_ct}) as
\begin{align}
&Y_{c}(t,f) = Z(c,t,f) + \sum\limits_{{c'=1, c'\neq c}}^C X_{c}(c',t,f) + \varepsilon_{c}(t,f) \nonumber \\
&= Z(c,t,f) + \sum\limits_{\substack{c'=1,\\c'\neq c}}^C {\mathbf{g}}_{c}(c',f)^{\H}\ \widetilde{{\mathbf{Z}}}(c',t,f) + \varepsilon'_{c}(t,f), \label{physical_model_ct_conv}
\end{align}
where, in the second row, $\widetilde{\mathbf{Z}}(c',t,f) = [Z(c',t-A+1,f), \mydots, Z(c',t,f), \mydots, Z(c',t+B,f)]^\T \in \CC^{A+B}$ stacks a window of $A+B$ T-F units, $\mathbf{g}_c(c',f) \in \CC^{A+B}$ is a sub-band filter, and $(\cdot)^\H$ computes Hermitian transpose.
In (\ref{physical_model_ct_conv}), we leverage narrow-band approximation \cite{Talmon2009CTF,Gannot2017,Wang2024USDnet} to approximate $X_c(c')$ (i.e., cross-talk speech of speaker $c'$ captured by the close-talk microphone of speaker $c$) as a linear convolution between a filter ${\mathbf{g}}_{c}(c',\cdot)$ and the close-talk speech $Z(c')$ of speaker $c'$, i.e., $X_c(c',t,f)\approx {\mathbf{g}}_{c}(c',f)^{\H}\ \widetilde{{\mathbf{Z}}}(c',t,f)$.
This is reasonable since $Z(c')$ can be roughly viewed as the dry source signal, and, in this case, we can interpret the filter as the acoustic transfer function from speaker $c'$ to close-talk microphone $c$.
In (\ref{physical_model_ct_conv}), $\varepsilon'$ contains the non-speech signal $\varepsilon$ and absorbs the modeling errors incurred by using narrow-band approximation.

Similarly, for each far-field mixture, we re-formulate (\ref{physical_model_ff}) as
\begin{align}
Y_p(t,f) &= \sum\nolimits_{c=1}^C {\mathbf{g}}_p(c,f)^{\H}\ \widetilde{{\mathbf{Z}}}(c,t,f) + \varepsilon'_p(t,f), \label{physical_model_ff_conv}
\end{align}
where ${\mathbf{g}}_p(c,f)$ can be interpreted as the acoustic transfer function from speaker $c$ to far-field microphone $p$.

Assuming that $\varepsilon'$ is weak, time-invariant and Gaussian, we can realize cross-talk reduction by solving, e.g., the problem below, which finds the sources, $Z(\cdot,\cdot,\cdot)$, and filters, $\mathbf{g}_{\cdot}(\cdot,\cdot)$, most consistent with the physical models in (\ref{physical_model_ct_conv}) and (\ref{physical_model_ff_conv}):
\begin{align}\label{ideal_loss}
&\underset{\mathbf{g}_{\cdot}(\cdot,\cdot),Z(\cdot,\cdot,\cdot)}{\text{argmin}}
\Big( \nonumber \\ 
&\sum\limits_{c=1}^C \sum\limits_{t,f} \Big| Y_c(t,f) - Z(c,t,f) - \sum\limits_{\substack{c'=1,\\c'\neq c}}^C {\mathbf{g}}_{c}(c',f)^{\H}\ \widetilde{{\mathbf{Z}}}(c',t,f) \Big|^2 \nonumber \\
&\quad\,\,\,\,\,+\sum\limits_{p=1}^{P} \sum\limits_{t,f} \Big| Y_p(t,f) - \sum\limits_{c=1}^C {\mathbf{g}}_p(c,f)^{\H}\ \widetilde{{\mathbf{Z}}}(c,t,f) \Big|^2 \Big),
\end{align}
where $|\cdot|$ computes magnitude.
This is a blind deconvolution problem \cite{Levin2011} in artificial intelligence and machine learning, which is highly non-convex and not solvable if no prior knowledge is assumed about the sources or the filters, since all of them are unknown and need to be estimated.

Motivated by the classic expectation-maximization algorithm in machine learning \cite{Bishop2006} and the recent UNSSOR algorithm \cite{Wang2023UNSSOR}, we propose to tackle this problem by leveraging unsupervised and weakly-supervised deep learning, where (a) a DNN is trained to first produce an estimate for each source (hence modeling source priors); (b) with the sources estimated, filter estimation in (\ref{ideal_loss}) then becomes a much simpler linear regression problem, where a closed-form solution exists and can be readily computed; and (c) loss functions similar to the objective in (\ref{ideal_loss}) can be designed to regularize the DNN estimates to have them respectively approximate the close-talk speech of each speaker.

\section{Unsupervised CTRnet}\label{proposed_algrithms_unsupervised}

Fig. \ref{system_figure} illustrates unsupervised CTRnet.
Given close-talk and far-field mixtures that are reasonably time-synchronized, it stacks close-talk and far-field mixtures as input, and produces, for each speaker $c$, an estimate $\hat{Z}(c)$, which, we will show, is regularized to be an estimate for close-talk speech $X_c(c)$.
Since $X_c(c)$ can be viewed as the dry source signal, we can reverberate its estimate $\hat{Z}(c)$ via a linear filtering algorithm named forward convolutive prediction (FCP) \cite{Wang2021FCPjournal} to identify speaker $c$'s cross-talk speech at the close-talk microphones of the other speakers, as well as its reverberant images at far-field microphones.
We realize this identification by training the DNN to optimize a combination of two loss functions, which encourage the DNN estimates and their linear filtering results to sum up to the mixture at each microphone (so that the linear filtering results can correctly identify the cross-talk speech which we aim at reducing).
At run time, $\hat{Z}(c)$ is used as the estimate of the close-talk speech $X_c(c)$ for each speaker $c$.
This section describes the DNN configurations, loss functions, and FCP filtering.

\subsection{DNN Configurations}

We stack the real and imaginary (RI) components of close-talk and far-field mixtures as input features for the DNN to predict $\hat{Z}(c)$ for each speaker $c$.
We can use complex spectral mapping \cite{Tan2020,Wang2020css}, where the DNN is trained to directly predict the RI components of $\hat{Z}(c)$ for each speaker $c$, or complex ratio masking \cite{Williamson2016}, where the DNN is trained to predict the RI components of a complex-valued mask $\hat{M}(c)$ for each speaker $c$ and $\hat{Z}(c)$ is then computed via $\hat{Z}(c) = \hat{M}(c) \odot Y_c$, with $\odot$ denoting point-wise multiplication.
The details of the DNN architecture are provided in Section \ref{system_config}, and the loss functions in the next subsection.

\subsection{Mixture-Constraint Loss}

We propose the following mixture-constraint (MC) loss:
\begin{align}\label{loss_MC}
\mathcal{L}_{\text{MC}} = \sum\nolimits_{c=1}^C \mathcal{L}_{\text{MC},c} + \alpha \times \sum\nolimits_{p=1}^P  \mathcal{L}_{\text{MC},p},
\end{align}
where $\mathcal{L}_{\text{MC},c}$ is the MC loss at close-talk microphone $c$, $\mathcal{L}_{\text{MC},p}$ at far-field microphone $p$, and $\alpha \in \RR_{>0}$ a weighting term.

Following the physical model in (\ref{physical_model_ct_conv}) and the first term in (\ref{ideal_loss}), at close-talk microphone $c$ we define $\mathcal{L}_{\text{MC},c}$ as
\begin{align}\label{loss_MC_close_talk_one}
&\mathcal{L}_{\text{MC},c} = \sum\nolimits_{t,f} \mathcal{F} \Big( Y_{c}(t,f), \hat{Y}_{c}(t,f) \Big) \nonumber \\
&= \sum\limits_{t,f} \mathcal{F} \Big( Y_{c}(t,f), \hat{X}_{c}(c,t,f) + \sum_{c'=1,c'\neq c}^C \hat{X}_{c}^{\text{FCP}}(c',t,f) \Big) \nonumber \\
&= \sum\limits_{t,f} \mathcal{F} \Big( Y_{c}(t,f), \hat{Z}(c,t,f) + \nonumber \\
&\quad\quad\quad\quad\quad\quad\quad\,\,\,\sum\nolimits_{c'=1,c'\neq c}^C \hat{\mathbf{g}}_{c}(c',f)^{\H}\ \widetilde{\hat{\mathbf{Z}}}(c',t,f) \Big),
\end{align}
where, in the third row, $\widetilde{\hat{\mathbf{Z}}}(c',t,f)=[\hat{Z}(c',t-I+1,f),\mydots,\hat{Z}(c',t,f),\mydots,\hat{Z}(c',t+J,f)]^\T \in \CC^{I+J}$ stacks a window of $I+J$ T-F units, and $\hat{\mathbf{g}}_{c}(c',f) \in \CC^{I+J}$ is a sub-band FCP filter to be described later in Section \ref{FCP_description}.
From row $2$ to $3$, we constrain the DNN estimate $\hat{Z}(c)$ of each speaker $c$ to be a gain- and time-aligned estimate of the close-talk speech (i.e., $\hat{X}_{c}(c)=\hat{Z}(c)$); and, for the reverberant image of each of the other speakers (i.e., cross-talk speech $X_{c}(c')$ with $c'\neq c$), we approximate it by linearly filtering $\hat{Z}(c')$ (i.e., $\hat{X}_{c}^{\text{FCP}}(c',t,f) = \hat{\mathbf{g}}_{c}(c',f)^{\H}\ \widetilde{\hat{\mathbf{Z}}}(c',t,f)$).
Since $\hat{Z}(c)$ is constrained to approximate $X_{c}(c)$, which can be viewed as the dry source signal, it is very reasonable to linearly filter $\hat{Z}(c)$ to estimate the image of speaker $c$ at the close-talk microphone of a different speaker and at far-field microphones.
The estimated speaker images are then added together to reconstruct the mixture: $\hat{Y}_{c}=\hat{X}_{c}(c) + \sum_{c'=1,c'\neq c}^C \hat{X}_{c}^{\text{FCP}}(c')$, which is reasonable under the assumption that the non-speech signal $\varepsilon$ is weak.
Finally, we compute a loss between the reconstructed mixture $\hat{Y}_c$ and the mixture $Y_c$, by using a function $\mathcal{F}(\cdot, \cdot)$, which computes an absolute loss on the estimated RI components and their magnitude \cite{Wang2022GridNetjournal}:
\begin{align}%
&\mathcal{F} \Big( Y_c(t,f), \hat{Y}_c(t,f) \Big) = \nonumber \\ &\quad\quad\quad\quad\quad\quad\quad\quad\quad\frac{\sum\nolimits_{\mathcal{O}\in \Omega} \Big| \mathcal{O}(Y_c(t,f)) - \mathcal{O}(\hat{Y}_c(t,f)) \Big| }{\sum\nolimits_{t',f'} \big| Y_c(t',f') \big|}, \nonumber
\end{align}
where $\Omega=\{\mathcal{R},\,\mathcal{I},\,\mathcal{A}\}$ is a set of functions with $\mathcal{R}(\cdot)$ extracting the real part, $\mathcal{I}(\cdot)$ the imaginary part and $\mathcal{A}(\cdot)$ the magnitude of a complex number, and the denominator balances the losses across different microphones and training mixtures.

Similarly, at far-field microphone $p$ we define $\mathcal{L}_{\text{MC},p}$ as
\begin{align}\label{loss_MC_far_field}
\mathcal{L}_{\text{MC},p} &= \sum\nolimits_{t,f} \mathcal{F} \Big( Y_p(t,f), \hat{Y}_p(t,f) \Big) \nonumber \\
&= \sum\nolimits_{t,f} \mathcal{F} \Big( Y_p(t,f), \sum\nolimits_{c=1}^C  \hat{X}_p^{\text{FCP}}(c,t,f) \Big) \nonumber \\
&= \sum\nolimits_{t,f} \mathcal{F} \Big( Y_p(t,f), \sum\limits_{c=1}^C \hat{\mathbf{g}}_p(c,f)^{\H}\ \widetilde{\hat{\mathbf{Z}}}(c,t,f) \Big),
\end{align}
where we linearly filter the DNN estimate $\hat{Z}(c)$ for each speaker $c$ using $\hat{\mathbf{g}}_p(c,f)$ so that their summation can approximate the mixture $Y_p$ captured by far-field microphone $p$.

\subsection{FCP for Filter Estimation}\label{FCP_description}

To compute $\mathcal{L}_{\text{MC}}$, we need to first estimate the linear filters, each of which is the relative transfer function relating the close-talk speech of a speaker to the speaker's reverberant image captured by a distant microphone.
Following UNSSOR \cite{Wang2023UNSSOR}, we employ FCP \cite{Wang2021FCPjournal} to estimate them.

Assuming speakers are non-moving within each utterance, we estimate the filters by solving the following problem:
\begin{align}\label{fcp_proj_mixture}
&\hat{\mathbf{g}}_r(c,f) = \nonumber \\
&\quad\quad\quad\,\,\,\,\underset{\mathbf{g}_r(c,f)}{\text{argmin}}
\sum\limits_t \frac{\Big| Y_r(t,f)-\mathbf{g}_r(c,f)^{\H}\ \widetilde{\hat{\mathbf{Z}}}(c,t,f) \Big|^2}{\hat{\lambda}_r(c,t,f)},
\end{align}
where symbol $r$ is used to index the $P$ far-field and $C$ close-talk microphones, and $\hat{\mathbf{g}}_r(c,f)$ and $\widetilde{\hat{\mathbf{Z}}}(c,t,f)$ are defined below (\ref{loss_MC_close_talk_one}).
$\hat{\lambda}$ is a weighting term balancing the importance of each T-F unit, and following \cite{Wang2021FCPjournal}, for each close-talk microphone $c$ and speaker $c'$, it is defined as
\begin{align}\label{FCPweight_ct}
\hat{\lambda}_c(c',t,f) = \xi\times \text{max}(|Y_c|^2) + |Y_c(t,f)|^2,
\end{align}
where $\xi$ floors the weighting term and $\text{max}(\cdot)$ extracts the maximum value of a power spectrogram; and for each far-field microphone $p$ and speaker $c$, it is defined as 
\begin{align}\label{FCPweight_ff}
\hat{\lambda}_p(c,t,f) = \xi\times \text{max}(|Y_p|^2) + Y_p(t,f).
\end{align}
We compute the weighting term individually for each microphone, considering that each speaker has different energy levels at different microphones.
(\ref{fcp_proj_mixture}) is a weighted linear regression problem, where a closed-form solution can be computed:
\begin{align}%
&\hat{\mathbf{g}}_r(c,f) = \nonumber \\
&\Big( \sum\limits_t \frac{\widetilde{\hat{\mathbf{Z}}}(c,t,f) \widetilde{\hat{\mathbf{Z}}}(c,t,f)^{\H}}{\hat{\lambda}_r(c,t,f)} \Big)^{-1} \sum\limits_t \frac{\widetilde{\hat{\mathbf{Z}}}(c,t,f) \big( Y_r(t,f) \big)^{*}}{\hat{\lambda}_r(c,t,f)},\nonumber
\end{align}
where $(\cdot)^{*}$ computes complex conjugate.
We then plug it into (\ref{loss_MC_close_talk_one}) and (\ref{loss_MC_far_field}), compute the losses, and train the DNN.

Although, in (\ref{fcp_proj_mixture}), $\hat{Z}(c)$ is filtered to approximate $Y_r$, earlier studies \cite{Wang2021FCPjournal} have suggested that the filtering result $\hat{\mathbf{g}}_r(c,f)^{\H}\ \widetilde{\hat{\mathbf{Z}}}(c,t,f)$ would approximate $X_r(c,t,f)$ (see the derivation in Appendix C of \cite{Wang2023UNSSOR}), if $\hat{Z}(c)$ is sufficiently accurate, which can be reasonably satisfied as the close-talk speech in close-talk mixtures already has a high input SNR.
The estimated speaker image is named \textit{FCP-estimated image} \cite{Wang2023UNSSOR}:
\begin{align}\label{FCP_image}
\hat{X}_r^{\text{FCP}}(c,t,f) = \hat{\mathbf{g}}_r(c,f)^{\H}\ \widetilde{\hat{\mathbf{Z}}}(c,t,f).
\end{align}
We can hence sum up the FCP-estimated images and compare the summation with $Y_r$ in (\ref{loss_MC_close_talk_one}) and (\ref{loss_MC_far_field}).
Notice that the FCP-estimated images at close-talk microphones represent the identified cross-talk speech we aim to reduce.

\section{Weakly-Supervised CTRnet}\label{weakly_supervised_CTRnet_description}

We propose to leverage speaker-activity timestamps as a weak supervision to improve the training of unsupervised CTRnet.
For each speaker $c$, its activity timestamps are denoted as a binary vector $d(c)\in \{0, 1\}^N$, where $N$ is signal length in samples, and value $1$ means that speaker $c$ is active at a sample and $0$ otherwise.
Note that speaker-activity timestamps are provided in almost every conversational speech recognition dataset.
They are annotated by having human annotators listen to close-talk mixtures.
This section describes why and how the timestamps can help training CTRnet. 

\subsection{Motivation}

In human conversations where concurrent speech naturally happens, speaker overlap is often sparse.
See Appendix \ref{illustration_sparse_overlap} for an example.
This sparsity poses major difficulties for speech separation \cite{Chen2020CSSdata,Cosentino2020LibriMix}, as the number of concurrent speakers is time-varying.
Modern separation models usually assume a maximum number of speakers the models can separate, but the models could over- or under-separate the speakers due to errors in speaker counting.

Assuming that there are at maximum $C$ speakers in each processing segment, unsupervised CTRnet, in essence, learns to produce $C$ output spectrograms that can be filtered to best \textit{explain} the recorded $C$ close-talk and $P$ far-field mixtures. 
One issue is that, in some training segments, if there are fewer than $C$ speakers, unsupervised CTRnet would always over-separate the speakers to obtain a smaller $\mathcal{L}_{\text{MC}}$ loss.\footnote{This is similar to unsupervised clustering. Hypothesizing more clusters for clustering can almost always produce a smaller objective but some clusters are split into smaller ones \cite{Bishop2006}.}
Therefore, the knowledge about the accurate number of speakers, which can be provided by speaker-activity timestamps, can be very useful to improve the training of unsupervised CTRnet.

Meanwhile, we find that unsupervised CTRnet usually cannot produce zero predictions in the silent ranges of each speaker.
In these ranges, the predicted spectrogram for each speaker often contains weak but non-negligible and intelligible signals from the other speakers.
This is a common problem in conversational speech separation \cite{Wang2020css,Morrone2023}.
In unsupervised CTRnet, this often results in a smaller $\mathcal{L}_{\text{MC}}$ loss, as the FCP filtering procedure could better approximate close-talk and far-field mixtures by filtering the non-zero predictions in silent regions.

With these understandings, we leverage speaker-activity timestamps in the following two ways to better train CTRnet.

\subsection{Muting during Training}

During training, we propose to mute all or part of the frames in each $\hat{Z}(c)$ based on the speaker-activity timestamps $d(c)$, before computing the filters and loss.
In detail, we compute 
\begin{align}\label{muting}
\hat{R}(c,t,f)=\hat{Z}(c,t,f) \times D(c,t) \times E(c),
\end{align}
where $D(c,t)\in \{0, 1\}$, defined based on $d(c)$, is $1$ if the STFT window corresponding to frame $t$ contains any active speech samples of speaker $c$ and is $0$ otherwise, and $E(c)\in \{0, 1\}$ is $1$ if $d(c)$ contains at least $0.1$ second of active speech in the training segment and is $0$ otherwise.
Using $D(c,t)$ for muting frames and $E(c)$ for muting speakers, we can avoid using non-zero predictions in silent ranges for FCP.
We then replace $\hat{Z}(c)$ with $\hat{R}(c)$ in (\ref{loss_MC_close_talk_one}), (\ref{loss_MC_far_field}) and (\ref{fcp_proj_mixture}) to compute FCP filters and $\mathcal{L}_{\text{MC}}$ for DNN training.
This way, CTRnet could more robustly reduce cross-talk speech for up to $C$ speakers.

\subsection{Speaker-Activity Loss}

We propose a speaker-activity (SA) loss $\mathcal{L}_{\text{SA},c}$ that can encourage the DNN-estimated signal for each speaker $c$ to be zero in silent ranges marked by speaker-activity timestamps:
\begin{align}\label{loss_SA}
\mathcal{L}_{\text{SA},c} = \frac{\big\| \hat{z}(c) \odot \big( 1-d(c) \big) \big\|_1}{\big\| y_c \odot \big( 1-d(c) \big) \big\|_1} \times \frac{N - \big\| d(c) \big\|_1}{N},
\end{align}
where $d(c)$ and $N$ are defined in the first paragraph of Section \ref{weakly_supervised_CTRnet_description}, $\|\cdot\|_1$ computes the $L_1$ norm, $\hat{z}(c)=\text{iSTFT}(\hat{Z}(c)) \in \RR^N$ is the re-synthesized time-domain signal of $\hat{Z}(c)$ obtained by applying inverse STFT (iSTFT) to $\hat{Z}(c)$, $y_c \in \RR^N$ is the time-domain close-talk mixture signal of speaker $c$, and the second term is a scaling factor accounting for the fact that each speaker usually has different length of silence in each training segment.
We combine it with $\mathcal{L}_{\text{MC}}$ in (\ref{loss_MC}) for training:
\begin{align}\label{loss_MC+SA}
\mathcal{L}_{\text{MC+SA}} = \mathcal{L}_{\text{MC}} + \beta \times \mathcal{L}_{\text{SA}} = \mathcal{L}_{\text{MC}} + \beta \times \sum\limits_{c=1}^C \mathcal{L}_{\text{SA},c},
\end{align}
where $\beta \in \RR_{>0}$ is a weighting term.
$\mathcal{L}_{\text{MC}}$ penalizes prediction errors in non-silent ranges of each speaker when muting is used, while $\mathcal{L}_{\text{SA}}$ penalizes predictions in silent ranges.

\section{Experimental Setup}\label{experimental_setup}

There are no existing studies on cross-talk reduction. 
We first evaluate unsupervised CTRnet on a simulated dataset named SMS-WSJ-FF-CT, where clean signals can be available for metric computation, and then evaluate weakly-supervised CTRnet on a real-recorded conversational automatic speech recognition (ASR) dataset named CHiME-7, using ASR metrics for evaluation.
This section describes the datasets, system setups, comparison systems, and evaluation metrics.

\subsection{SMS-WSJ-FF-CT and Evaluation Setup}

SMS-WSJ-FF-CT, with ``FF'' meaning \textit{far-field} and ``CT'' \textit{close-talk}, is built upon a simulated dataset named SMS-WSJ \cite{Drude2019} which consists of $2$-speaker fully-overlapped noisy-reverberant mixtures, by simulating a close-talk microphone for each speaker.
Fig. \ref{physical_model_figure} shows this setup.
This evaluation serves as a proof of concept to validate whether CTRnet can work well in ideal cases, where the hypothesized physical models in (\ref{physical_model_ct_conv}) and (\ref{physical_model_ff_conv}) are largely satisfied.

\textbf{SMS-WSJ} \cite{Drude2019}, a popular corpus for evaluating $2$-speaker separation algorithms in reverberant conditions, has $33,561$ ($\sim$$87.4$ h), $982$ ($\sim$$2.5$ h) and $1,332$ ($\sim$$3.4$ h) $2$-speaker mixtures for training, validation and testing.
The clean speech is sampled from the WSJ0 and WSJ1 corpus.
The simulated far-field microphone array has $6$ microphones uniformly placed on a circle with a diameter of $20$ cm.
For each mixture, the speaker-to-array distance is sampled from the range $[1.0, 2.0]$ m, and the reverberation time (T60) from $[0.2, 0.5]$ s.
A weak white noise is added to simulate microphone self-noises, at an energy level between the summation of the reverberant speech signals and the noise sampled from the range $[20, 30]$ dB.
The sampling rate is 8 kHz.

\textbf{SMS-WSJ-FF-CT} is simulated by adding a close-talk microphone for each of the speakers in each SMS-WSJ mixture.
In each mixture, the distance from each speaker to its close-talk microphone is uniformly sampled from the range $[0.1, 0.3]$ m, and the angle from $[-\pi, \pi]$.
All the other setup remains unchanged.
This way, we can obtain the close-talk mixture of each speaker, and the far-field mixtures are exactly the same as those in SMS-WSJ.

\textbf{Training and Inference of CTRnet}.
For training, in default we sample an $L$-second segment from each mixture in each epoch, and the batch size is $H$.
For inference, we feed each test mixture in its full length to CTRnet.

\textbf{Evaluation Metrics}. For each speaker $c$, we use its time-domain close-talk speech corresponding to $X_c(c)$ as the reference signal for evaluation.
The evaluation metrics include signal-to-distortion ratio (SDR) \cite{Vincent2006a}, scale-invariant SDR (SI-SDR) \cite{LeRoux2019}, perceptual evaluation of speech quality (PESQ) \cite{Rix2001}, and extended short-time objective intelligibility (eSTOI) \cite{H.Taal2011}. All of them are widely-used in speech separation.

\textbf{Comparison Systems} include signal processing based unsupervised speech separation algorithms, cACGMM-based spatial clustering (SC) \cite{Boeddeker2021ITG} and independent vector analysis (IVA) \cite{Sawada2019BSSReview}.
Both are very popular.
We stack close-talk and far-field mixtures as input.
For SC, we use a public implementation \cite{Boeddeker2019SpatialClustering} provided in the \textit{pb\_bss} toolkit; and for IVA, we use the \textit{torchiva} toolkit \cite{Scheibler2022}.
The STFT window and hop sizes are tuned to 128 and 16 ms for SC, and to 256 and 32 ms for IVA.
For IVA, we use the Gaussian model in \textit{torchiva} to model source distribution.
A garbage source is used in both models to absorb modeling errors.

\subsection{CHiME-7 and Evaluation Setup}\label{chime7_setup}

To show that CTRnet can work on realistic data, we train and evaluate CTRnet using the real-recorded CHiME-7 dataset, following the setup of the CHiME-7 DASR challenge \cite{cornell23_chime}.
CHiME-7, built upon CHiME-\{5,6\} \cite{Watanabe2020CHiME6,Barker2018CHiME5}, is a notoriously difficult dataset in conversational speech separation and recognition, mainly due to its realisticness, which is representative of common problems deployed systems could run into in practice, such as microphone synchronization errors, signal clipping, frame dropping, microphone failures, moving arrays, moving speakers, varying degrees of speaker overlap, and challenging environmental noises.
So far, the most successful separation algorithm for CHiME-7 is still based on guided source separation (GSS) \cite{Boeddecker2018GSS}, a signal processing algorithm, and supervised DNN-based approaches have nearly no success on this dataset.
All the top teams in CHiME-7 \cite{Wang2023CHiME7winner,Ye2023CHiME7} and in similar challenges or datasets, e.g., AliMeeting \cite{Yu2022M2MeT,Liang2023M2MeT2} and AMI \cite{Raj2022GPUGSS}, all adopt GSS as the only separation module.
This section describes the dataset and our evaluation setup.

\textbf{CHiME-7 Dataset} contains real-recorded conversational sessions, each with $4$ participants speaking spontaneously in a domestic, dinner-party scenario, where concurrent speech can naturally happen.
Each speaker (participant) wears a binaural close-talk microphone, and there are $6$ Kinect devices, each with $4$ microphones, placed in a strategic way in the room to record each entire session, which is $1.5$ to $2.5$ hours long.
The recorded close-talk mixtures contain severe cross-talk speech.
Realistic noises typical in dinner parties are recorded at the same time along with speech.
There are 14 ($\sim$$34$ h), 2 ($\sim$$2$ h) and 4 ($\sim$$5$ h) recorded sessions respectively for training, validation and testing.
The sampling rate is $16$ kHz.

\textbf{Experiment Design}.
Based on the CHiME-7 dataset, we design an experiment for cross-talk reduction.
We consider the mixture recorded at the right ear of each binaural microphone as the close-talk mixture and the one at the left ear as far-field mixture, meaning that we have $C=4$ close-talk and $P=4$ far-field mixtures.

\textbf{Training and Inference of CTRnet}.
To train CTRnet, we cut each (long) session into $8$-second segments with $50\%$ overlap between consecutive segments, and train CTRnet on these segments.
At inference time, we apply the trained CTRnet in a block-wise way to process each (long) session.
The final separated speech for each speaker has the same length as the (long) input session.
See Appendix \ref{sec:app:block_processing} for the details.

\textbf{Evaluation Metrics}.
With the separated signal of each speaker, following the CHiME-7 DASR challenge \cite{cornell23_chime} we compute diarization-assigned word error rates (DA-WER).
In detail, the separated signal of each speaker is split to short utterances by using oracle speaker-activity timestamps, and the default, pre-trained ASR model provided by the challenge is used to recognize and score each utterance.
The pre-trained ASR model\footnote{\url{https://huggingface.co/popcornell/chime7_task1_asr1_baseline}} is a strong end-to-end model based on a transformer encoder/decoder architecture trained with joint CTC/attention, using WavLM features. 

\begin{table*}[]
\scriptsize
\centering
\sisetup{table-format=2.2,round-mode=places,round-precision=2,table-number-alignment = center,detect-weight=true,detect-inline-weight=math}
\setlength{\tabcolsep}{2pt}
\resizebox{1.785\columnwidth}{!}{
\begin{tabular}{
r %
c %
S[table-format=2,round-precision=0] %
S[table-format=1,round-precision=0] %
S[table-format=1,round-precision=0] %
S[table-format=1,round-precision=0] %
c %
c %
c %
S[table-format=2.1,round-precision=1]
S[table-format=2.1,round-precision=1]
S[table-format=1.2,round-precision=2]
S[table-format=1.3,round-precision=3]
}
\toprule
Row & Systems & {$I$} & {$J$} & {$C$} & {$P$} & {Masking/Mapping} & {$\alpha$} & {$H/L$} & {SI-SDR (dB) $\uparrow$} & {SDR (dB) $\uparrow$} & {PESQ $\uparrow$} & {eSTOI $\uparrow$} \\

\midrule
0 & Unprocessed mixture & {-} & {-} & {-} & {-} & {-} & {-} & {-} & 14.655501344786572 & 14.70267880020375 & 2.9178258743998526 & 0.8748214897000004 \\
\midrule
1a & Unsupervised CTRnet & 20 & 0 & 2 & 6 & {Mapping} & $1.0$ & {$4\,/\,4$} & 23.974319848626315 & 24.31962710414399 & 3.7712965533808545 & 0.9640058551369834 \\
1b & Unsupervised CTRnet & 30 & 0 & 2 & 6 & {Mapping} & $1.0$ & {$4\,/\,4$} & 24.12571720778942 & 24.464936434723874 & 3.813656710334368 & 0.9650061299316992 \\
1c & Unsupervised CTRnet & 40 & 0 & 2 & 6 & {Mapping} & $1.0$ & {$4\,/\,4$} & 23.892514861269934 & 24.247414572603518 & 3.8577268699088973 & 0.9646202934468251 \\
1d & Unsupervised CTRnet & 50 & 0 & 2 & 6 & {Mapping} & $1.0$ & {$4\,/\,4$} & 23.361094367911328 & 23.7223341860702 & 3.8644242688968733 & 0.9624741013662066 \\
\midrule
2 & Unsupervised CTRnet & 30 & 0 & 2 & 6 & {Masking} & $1.0$ & {$4\,/\,4$} & 23.380081465428656 & 23.730775207059608 & 3.9009404569506287 & 0.9644549687904143 \\
\midrule
3a & Unsupervised CTRnet & 30 & 0 & 2 & 6 & {Mapping} & $C/P$ & {$4\,/\,4$} & 25.843364990598808 & 26.127218565852772 & \bfseries 3.9082714844573365 & 0.9713115427474723 \\
3b & Unsupervised CTRnet & 30 & 0 & 2 & 6 & {Mapping} & $1/P$ & {$4\,/\,4$} & 26.00108980545187 & 26.281085589902595 & 3.8975711828685022 & 0.9711379708297556 \\
3c & Unsupervised CTRnet & 30 & 0 & 2 & 6 & {Mapping} & $1/(2\times P)$ & {$4\,/\,4$} & 25.94988372076202 & 26.226075532525165 & 3.898657554620737 & 0.9709378615139035 \\
3d & Unsupervised CTRnet & 30 & 0 & 2 & 6 & {Mapping} & $1/(3\times P)$ & {$4\,/\,4$} & 25.89058857205584 & 26.162292488269404 & 3.873716520788791 & 0.9701841274639627 \\
\midrule
4a & Unsupervised CTRnet & 29 & 1 & 2 & 6 & {Mapping} & $1/P$ & {$4\,/\,4$} & 25.87792800498886 & 26.14939767077047 & 3.89687808573962 & 0.9706188397272134 \\
4b & Unsupervised CTRnet & 28 & 2 & 2 & 6 & {Mapping} & $1/P$ & {$4\,/\,4$} & 25.723054752246036 & 26.00193140619539 & 3.903394768993418 & 0.9705644009879817 \\
4c & Unsupervised CTRnet & 27 & 3 & 2 & 6 & {Mapping} & $1/P$ & {$4\,/\,4$} & 25.918182888859743 & 26.18723605026406 & 3.883531636035478 & 0.9707010214724079 \\
\midrule
5a & Unsupervised CTRnet & 30 & 0 & 2 & 3 & {Mapping} & $1/P$ & {$4\,/\,4$} & 25.89469175718658 & 26.16200267889599 & 3.8736722487229125 & 0.9708929735599938 \\
5b & Unsupervised CTRnet & 30 & 0 & 2 & 2 & {Mapping} & $1/P$ & {$4\,/\,4$} & 25.894458302416957 & 26.15790520553707 & 3.8745971420058263 & 0.9703828428443109 \\
5c & Unsupervised CTRnet & 30 & 0 & 2 & 1 & {Mapping} & $1/P$ & {$4\,/\,4$} & 24.754759183561838 & 25.01522663407262 & 3.8666828817075434 & 0.9680884638973974 \\
\midrule
6 & Unsupervised CTRnet & 30 & 0 & 2 & 6 & {Mapping} & $1/P$ & {$1\,/\,-$} & \bfseries 26.46133595691608 & \bfseries 26.75524646158299 & 3.878209313875562 & \bfseries 0.9731036914117346 \\
\midrule
7a & SC \cite{Boeddeker2019SpatialClustering} & {-} & {-} & {-} & 6 & {-} & {-} & {-} & -1.9298528639103818 & 7.069643522805861 & 2.266987637043357 & 0.5611181169390558 \\
7b & IVA \cite{Scheibler2022} & {-} & {-} & {-} & 6 & {-} & {-} & {-} & 22.630424480226715 & 23.69221788104909 & 3.659791785049009 & 0.9476332602765591 \\
\bottomrule
\end{tabular}
}
\caption{Averaged separation results of unsupervised CTRnet on SMS-WSJ-FF-CT.}\label{ctr_results}
\end{table*}

\textbf{Comparison Systems}.
We use GSS \cite{Boeddecker2018GSS} for comparison, by following the implementation provided in CHiME-7 DASR Challenge.\footnote{See \href{https://github.com/espnet/espnet/blob/master/egs2/chime7_task1/asr1/local/run_gss.sh}{https://github.com/espnet/espnet/blob/master/egs2/chime7\_ task1/asr1/local/run\_gss.sh}}
GSS is the most popular and effective separation model so far for modern ASR systems.
It first performs dereverberation using the weighted prediction algorithm \cite{Nakatani2010} and then computes a mask-based beamformer for separation by using posterior time-frequency masks estimated by a spatial clustering module guided by oracle speaker-activity timestamps \cite{Boeddecker2018GSS}.
Notice that at run time GSS requires oracle speaker-activity timestamps, while weakly-supervised CTRnet only needs them for training and, once trained, no longer needs them.
Another note is that the training data of the pre-trained ASR model provided by the challenge contains GSS-processed signals, and is hence favorable to GSS.

\subsection{Miscellaneous Configurations of CTRnet}\label{system_config}

For STFT, the window size is $16$ ms, hop size $8$ ms, and the square root of the Hann window is used as the analysis window.
TF-GridNet \cite{Wang2022GridNetjournal} is employed as the DNN architecture.
Using the symbols defined in Table I of \cite{Wang2022GridNetjournal}, we set its hyper-parameters to $D=128$, $B=4$, $I=1$, $J=1$, $H=192$, $L=4$ and $E=4$ (please do not confuse these symbols with the ones defined in this paper).
The model has around $4.8$ million parameters.
$\xi$ in (\ref{FCPweight_ct}) and (\ref{FCPweight_ff}) is tuned to $10^{-3}$.
$\beta$ in (\ref{loss_MC+SA}) is set to $1.0$.

\section{Evaluation Results and Discussions}

\subsection{Results on SMS-WSJ-FF-CT}

Table \ref{ctr_results} configures unsupervised CTRnet in various ways and presents the results on SMS-WSJ-FF-CT.
Row $0$ reports the scores of unprocessed mixtures.
The $14.7$ dB SI-SDR indicates that the close-talk mixtures are not clean due to the contamination by cross-talk speech.
In 1a-1d, we vary the number of FCP filter taps $I+J\in \{20, 30, 40, 50\}$ and configure the FCP filters to be causal by setting $J=0$.
We observe that the setup in 1b obtains the best SI-SDR.
In row $2$, complex masking rather than mapping is used to obtain $\hat{Z}(c)$, but the result is not better than 1b.
In 3a-3d, we reduce $\alpha$ in (\ref{loss_MC}) from $1.0$ to $C/P, 1/P, 1/(2\times P)$ and $1/(3\times P)$ so that the loss on close-talk mixtures is emphasized.
This is reasonable since we aim at separating close-talk speech, rather than far-field speaker images.
From 1b and 3b, we observe that this change leads to clear improvement (e.g., from $24.1$ to $26.0$ dB SI-SDR).
In 4a-4c, we use non-causal FCP filters by increasing $J$ from $0$ to $1$, $2$ and $3$, while fixing the total number of filter taps to $30$.
This does not yield improvements, likely because $\hat{Z}(c)$ is regularized to be an estimate of the close-talk speech of speaker $c$ and hence the transfer function relating it to speaker $c$'s reverberant images at other microphones should be largely causal.
In 5a-5c, we reduce the number of far-field microphones from $6$ to $3$, $2$ and $1$.
That is, we only use microphone (a) $1$; (b) $1$ and $4$; and (c) $1$, $3$ and $5$ of the far-field six-microphone array to simulate the cases when the far-field array only has a limited number of microphones.
Compared with 3b, the performance drops, indicating the benefits of using more far-field mixtures as network input and for loss computation, while the improvement over the mixture scores in row $0$ is still large.
In default, our models are trained using mini-batches of 4 four-second segments, while, in row $6$, we train the model by using a batch size of $1$ and using each training mixture in its full length.
This improves SI-SDR over 3b, likely because better FCP filters can be computed during training by using all the frames in each mixture.

In row 7a and 7b, SC and IVA perform worse than CTRnet.
SC does not work well, possibly because, in this distributed-microphone scenario where each speaker signal can have very different SNRs at different microphones, the target T-F masks at different microphones are significantly different.

\begin{table}[]
\scriptsize
\centering
\sisetup{table-format=2.2,round-mode=places,round-precision=2,table-number-alignment = center,detect-weight=true,detect-inline-weight=math}
\setlength{\tabcolsep}{1.5pt}
\resizebox{1.0\columnwidth}{!}{
\begin{tabular}{
r %
c %
c %
S [table-format=2,round-precision=0] %
S [table-format=1,round-precision=0] %
c %
c %
S[table-format=2.1,round-precision=1] %
S[table-format=2.1,round-precision=1] %
}
\toprule
 & & & & & & & \multicolumn{2}{c}{DA-WER (\%) $\downarrow$} \\
\cmidrule(lr{9pt}){8-9}
Row & Systems & Muting? & $I$ & $J$ & $C$ & $P$ & \multicolumn{1}{c}{Val.} & \multicolumn{1}{c}{Test} \\

\midrule
0 & Unprocessed mixture & - & {-} & {-} & 4 & {-} & 28.2791 & 27.7891 \\
\midrule
1 & Unsupervised CTRnet & - & 19 & 1 & $4$ & $4$ & 22.5013 & 25.1378 \\
2 & Weakly-supervised CTRnet & \xmark & 19 & 1 & $4$ & $4$ & 79.1359 & 72.9919 \\
3 & Weakly-supervised CTRnet & \cmark & 19 & 1 & $4$ & $4$ & 20.5143 & 22.5746 \\
\midrule
4 & GSS \cite{Boeddecker2018GSS} & - & {-} & {-} & $4$ & $4$ & 26.2343 & 26.6265 \\
\bottomrule
\end{tabular}
}
\caption{ASR results of CTRnet on CHiME-7 close-talk mixtures.}\label{ctr_results_CHiME7}
\end{table}

\subsection{Results on CHiME-7}

Table \ref{ctr_results_CHiME7} reports the ASR results of CTRnet on the close-talk mixtures of CHiME-7.
The filter taps $I$ and $J$ are tuned to $19$ and $1$.
Notice that, here, one future tap is used, as the real-recorded data in CHiME-7 exhibits non-negligible synchronization errors among different microphone signals and we find that allowing one future tap can mitigate the synchronization issues.
Complex spectral mapping is used in default.

In row $0$, the mixture DA-WER is high even though a strong pre-trained ASR model is used for recognition.
This is because the close-talk mixtures contain very strong cross-talk speech, which confuses the ASR model on which speaker to recognize.
In row $1$, using unsupervised CTRnet to reduce cross-talk speech produces clear improvement (from $27.8\%$ to $25.1\%$ DA-WER).
In row $3$, weakly-supervised CTRnet with muting during training further improves the performance to $22.6\%$.
In row $2$, muting is not applied when training weakly-supervised CTRnet, and we observe much worse DA-WER.
We found that this is because, without using muting, the $\mathcal{L}_{\text{SA}}$ loss in (\ref{loss_MC+SA}) tends to push $\hat{Z}(c)$ of each speaker $c$ towards zeros to obtain a smaller loss value.
Compared with GSS, weakly-supervised CTRnet, through learning, obtains clearly better DA-WER (i.e., $22.6\%$ vs. $26.6\%$).
In addition, clear improvement is obtained over the unprocessed mixtures (i.e., $22.6\%$ vs. $27.8\%$).
These results indicate the effectiveness of CTRnet for cross-talk reduction on real-recorded data.

\section{Conclusion}

We have proposed a novel task, cross-talk reduction, and a novel solution, CTRnet, with or without leveraging speaker-activity timestamps for model training.
A key contribution of this paper, we emphasize, is that the proposed CTRnet can be trained directly on real-recorded pairs of far-field and close-talk mixtures, and is capable of effectively reducing cross-talk speech, especially on the notoriously difficult real-recorded CHiME-7 data.
This contribution suggests a promising way towards addressing a fundamental limitation of real-recorded close-talk speech.
That is, the contamination by cross-talk speech, which makes close-talk mixtures not sufficiently clean.
With CTRnet producing reasonably-good cross-talk reduction, we expect many applications to be enabled, and we will investigate them in our future work.

In closing, another key contribution of this paper, we point out, is that the proposed weakly-supervised deep learning based methodology for blind deconvolution can work well on challenging real-recorded data.
This contribution, we believe, would motivate a new stream of research towards realizing robust neural speech separation in realistic conditions, and generate broader impact beyond CTR and speech separation, especially in many machine learning and artificial intelligence applications where the sensors would not only capture target signals but also interference signals very detrimental to machine perception.

\appendix

\section{Run-Time Separation of Long  Sessions}
\label{sec:app:block_processing}

In CHiME-$7$, the duration of each recorded session ranges from $1.5$ to $2.5$ hours.
At run time, to separate the close-talk speech of an entire session, we run CTRNet in a block-wise way, using the pseudo-code below at each processing block:
\begin{align}
Y_c &:= Y_c / \sigma_c, \text{for}\,c\in\{1,\mydots,C\}, \\ 
Y_p &:= Y_p / \sigma_p, \text{for}\,p\in\{1,\mydots,P\}, \\
\hat{Z}(1),\mydots,\hat{Z}(C) &= \texttt{CTRnet}(Y_1,\mydots,Y_C,Y_1,\mydots,Y_P), \label{dummytmptmp}\\
\hat{Z}(c) &:= \hat{Z}(c) \times \sigma_c, \text{for}\,c\in\{1,\mydots,C\},
\label{block_online_gain}
\end{align}
where $\sigma_c$ and $\sigma_p\in \RR$ are respectively the sample-level standard deviations of the time-domain close-talk mixture $y_c$ and far-field mixture $y_p$.
After obtaining $\hat{Z}(c)$ at each block, we stack $\hat{Z}(c)$ of all the processing blocks and then revert the stacked spectrogram to time domain via inverse STFT.
Notice that $\mathcal{L}_{\text{MC},c}$ in (\ref{loss_MC_close_talk_one}) constrains each close-talk speech estimate $\hat{Z}(c)$ to have the same gain level as the close-talk mixture $Y_c$, by not filtering $\hat{Z}(c)$.
As a result, in (\ref{dummytmptmp}), each output $\hat{Z}(c)$ is expected to have the same gain level as $Y_c$.

For our experiments on CHiME-$7$, the processing block size is set to $8$ seconds, the same as the segment length used during training.
We configure the blocks to be slightly overlapped, where we consider the first and the last $0.96$ seconds as context, and output the DNN estimates in the center $6.08$ ($=8-0.96-0.96$) seconds at each block.

\section{Illustration of Sparse Speaker Overlap}
\label{illustration_sparse_overlap}

\begin{figure}[H]
  \centering  
  \includegraphics[width=8cm]{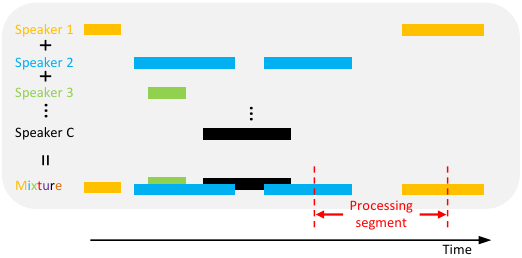}
  \caption{
   Illustration of sparse speaker overlap in human conversations.
   Best viewed in color.
   Each colored band means that the corresponding speaker is talking in the time range.
  }
  \label{sparse_overlap_figure}
\end{figure}

Fig. \ref{sparse_overlap_figure} illustrates sparse speaker overlap in human conversations.
This sparsity is because, for conversations in, e.g., meeting or dinner-party scenarios, people often take turns to speak and tend to not always speak at the same time.
As a result, in different (short) processing segments of separation systems, the number of active speakers and the degrees of speaker overlap can vary a lot.

\clearpage

\section*{Acknowledgments}

Experiments of this work used the Bridges2 system at PSC and Delta at NCSA through allocation CIS$210014$ and IRI$120008$P from the Advanced Cyberinfrastructure Coordination Ecosystem: Services and Support (ACCESS) program, which is supported by National Science Foundation grants \#$2138259$, \#$2138286$, \#$2138307$, \#$2137603$ and \#$2138296$.

\bibliographystyle{named}
\bibliography{ijcai24}

\end{document}